\documentclass[preprint2]{aastex}

\usepackage{natbib}
\usepackage{graphicx}
\usepackage{multirow}
\usepackage{txfonts}

\begin{document}

\shorttitle{Accretion Flow Properties of Swift~J1753-0127}
\shortauthors{D. Debnath, A. Jana, S. K. Chakrabarti et al.}

\title{Accretion Flow Properties of Swift~J1753.5-0127 during its 2005 outburst}
\author{Dipak Debnath\altaffilmark{1}, Arghajit Jana\altaffilmark{1}, Sandip K. Chakrabarti\altaffilmark{2,1}, 
Debjit Chatterjee\altaffilmark{1}, Santanu Mondal\altaffilmark{3,1} }
\altaffiltext{1}{Indian Center for Space Physics, 43 Chalantika, Garia St. Rd., Kolkata, 700084, India.}
\altaffiltext{2}{S. N. Bose National Centre for Basic Sciences, Salt Lake, Kolkata, 700098, India.}
\altaffiltext{3}{Instituto de F\'isica y Astronom\'ia, Facultad de Ciencias, Universidad de Valpara\'iso, Gran Bretana N 1111, Playa Ancha, Valparaíso, Chile}

\email{dipak@csp.res.in; argha@csp.res.in; chakraba@bose.res.in; debjit@csp.res.in; santanu@csp.res.in}


\begin{abstract}

Galactic X-ray binary black hole candidate Swift~J1753.5-0127 was discovered on June 30 2005 by Swift/BAT instrument. 
In this paper, we make detailed analysis of spectral and timing properties of its 2005 outburst using RXTE/PCA 
archival data. We study evolution of spectral properties of the source from spectral analysis with the additive table model 
{\it fits} file of the Chakrabarti-Titarchuk two-components advective flow (TCAF) solution. From spectral fit, we extract 
physical flow parameters, such as, Keplerian disk accretion rate, sub-Keplerian halo rate, shock location and shock compression 
ratio, etc. We also study the evolution of temporal properties, such as observation of low frequency quasi-periodic oscillations (QPOs), 
variation of X-ray intensity throughout the outburst. From the nature of the variation of QPOs, and accretion rate ratios 
(ARRs=ratio of halo to disk rates), we classify entire 2005 outburst into two harder (hard-intermediate and hard) 
spectral states. No signature of softer (soft-intermediate and soft) spectral states are seen. This may be because of significant halo rate 
throughout the outburst. This behavior is similar to a class of other short orbital period sources, such as, MAXI~J1836-194, MAXI~J1659-152 
and XTE~J1118+480. Here, we also estimate probable mass range of the source, to be in between $4.75 M_\odot$ to $5.90 M_\odot$ 
based on our spectral analysis.

\end{abstract}


\section{Introduction}

Galactic black holes candidates (BHCs) are situated generally in binary systems. Some of these sources are transient in nature.
These black hole X-ray binaries (BHXRB) spend most of their lifetime in the quiescent states. Occasionally, they show 
outbursts in which high energy radiation intensity increases by a factor of hundred or more. A possible explanation 
is that an outburst is triggered by the sudden rise in viscosity in the accreting matter coming from the companion 
(Chakrabarti \& Titarchuk 1995; Ebisawa et al. 1996). The mass transfer from the companion to the compact object 
can occur via Roche lobe overflow or by wind accretion depending on the nature of the companion. 
In general, if the companion is a low mass K or M type star, Roche lobe overflow is expected. If the companion 
is a mass losing high massive star, such as a type of O, A or B, winds form and accretion of winds are expected. 
In a BHXRB, the matter does not fall radially on the compact object. Rather, it forms an inward 
spiraling accretion disk around it. This disk emits electromagnetic 
radiation in the range from radio to $\gamma$-rays.
High energy radiations, such as X-rays, are very important to study the properties of black holes (BHs), 
as they come from regions close to the BH event horizon and therefore carry accurate information about the compact object. 

In general, X-ray spectrum consists of a multi-color black body type component of soft photons and a power-law (PL) type tail 
of hard photons. The multi-color black body component is believed to come from the standard accretion disk (Shakura \& Sunyaev 1973;
Novikov \& Throne 1973) while the PL component is believed to come from a hot-electron 
filled Compton cloud region (Sunyaev \& Titarchuk 1980, 1985).
In the two component advective flow (TCAF) solution Chakrabarti and his collaborators (Chakrabarti \& Titarchuk, 1995, hereafter CT95; 
Chakrabarti, 1997), the so-called Compton cloud is replaced with CENtrifugal pressure supported BOundary Layer (CENBOL). 
In the TCAF solution, the accretion disk consists of two components of accreting matter -  a standard Keplerian disk 
with temperature modified by the Compton cloud and a sub-Keplerian halo.
The Keplerian disk is a high viscous disk with high angular momentum while the sub-Keplerian disk has low viscosity 
and low angular momentum. The sub-Keplerian halo piles up at the centrifugal barrier and an axisymmetric shock is formed 
(Chakrabarti 1990). The soft photons from the Keplerian disk are inverse-Comptonized at the post-shock region 
and become hard photons. CENBOL is also considered to be the base of jet (Chakrabarti 1999). 
Its oscillation is believed to cause low frequency quasi-Periodic Oscillations (QPOs). QPOs are observed if resonance conditions 
are satisfied when infall timescale roughly matches with the cooling time scale (Molteni et al. 1996; Chakrabarti et al. 2015) or 
the Rankine-Hugoniot conditions are not satisfied (Ryu et al. 1997). The observed frequency of the QPOs are inversely 
proportional to the infall time in the post-shock region (Chakrabarti \& Manickam 2000).

Recently, Debnath and his collaborators' have successfully included this most generalized accretion solution 
into HEASARC's spectral analysis software package XSPEC (Arnaud 1996) as an additive table model (Debnath et al. 2014; 
Mondal et al. 2014, hereafter MDC14; Debnath et al. 2015a,b, hereafter DMC15, DMCM15 respectively; Jana et al. 2016, hereafter JDCMM16; 
Chatterjee et al. 2016, hereafter CDCMJ16) to find the nature of accretion flow properties of a few transient BH sources 
during their X-ray outbursts. Apart from the mass of the black hole and constant normalization, one requires four parameters 
(two types of accretion rates, shock location and compression ratio) to fit a BH spectrum with the TCAF solution in XSPEC. 
Transient BHXRBs show several spectral states during an entire outburst, which could be well understood by observing variations 
of these physical flow parameters and nature of QPOs. Four spectral states, namely, hard (HS), hard-intermediate (HIMS), 
soft intermediate (SIMS) and soft state (SS) are commonly observed in an outburst of a transient BHC. 
In a typical outburst, BHCs generally stay in the HS at the start of an outburst. After that, it enters into the HIMS, SIMS and SS 
one by one. The shock moves closer to the black hole as it moves to the softer state and before disappearing altogether 
in soft/very soft state. The declining phase starts when the matter supply to the inner disk is turned off. During this phase 
of an outburst, the shock moves away from the black hole and the spectra start to become harder. Generally SIMS, HIMS and HS 
are observed one after another, to complete the hysteresis loop of the spectral states (see, Debnath et al. 2013 and references therein). 

In general, in the harder states (HS and HIMS), the halo rate dominates while in the softer states (SIMS and SS), the disk rate dominates.
From the spectral fits of data of each day with TCAF solution, we can also estimate the mass of black hole more accurately using 
constant normalization method  (Molla et al. 2016; hereafter MDCMJ16). 
According to the TCAF solution, model normalization (N) should not vary on a daily basis, since it only depends on the mass, distance 
and disk inclination angle $`i'$ of the disk, unless precession in the disk occurs to change the projected surface of the emission area 
or there are some significant outflow activities which are not included in current TCAF model {\it fits} file. 
This method has been successfully applied to estimate masses of few transient Galactic BHCs, such as, MAXI~J1659-152 (MDCMJ16), 
MAXI~J1836-194 (JDCMM16), MAXI~J1543-564 (CDCMJ16), H~1743-322 (Bhattacharjee et al. 2017; Molla et al. 2017). 
This motivated us to estimate the mass of Swift~J1753.5-0127 using the same method.
Moreover, Mondal et al. (2016) also estimated spin parameter of BHC GX~339-4, from their spectral study of the relativistic broadening 
of the double Fe-line with combined TCAF plus LAOR model fits using Swift/XRT and NuSTAR/FMA data of the 2013 outburst.

Swift~J1753.5-0127 was discovered by Swift/BAT on 2005 June 30 (Palmar et al. 2005) at the sky position 
of RA$=17^h 53^m 28^s.3 $, DEC$=-01^\circ 27' 09''.3$. It was also detected in optical (Halpern et. al 2005) 
and radio (Fender et al. 2005). BHC Swift~J1753.5-0127 has a short orbital period of $3.2\pm 0.2 $ hrs 
(Zurita et al. 2007). However Neustroev et al. (2014) suggested the orbital period is $2.85$~hrs.
They also reported that the binary has a primary BH mass of less than $5~M_{\odot}$ and companion 
as $0.17-0.25 M_{\odot}$ and disk inclination angle of $>40^\circ$.
Recently, Shaw et al. (2016) predicted the mass of central BH to be greater than $7.4~M_{\odot}$.
Cadolle Bel et al. (2007) estimated the distance of the system is likely to be in between $4-8$~kpc.
However, from UV spectral study, Froning et al. (2014) calculated the source distance as $< 3.7$~kpc
for inclination of $i=55^\circ$ and $< 2.8$ kpc for inclination of $i=0^\circ$.
The secondary companion is also reported to be a K or M type main-sequence star (Bell et. al 2007).
Radio jet was also observed during the 2005 outburst of source (Fender et al. 2005; Soleri et al. 2010).
The source is still active in HS with observable low luminosity X-ray intensities. In fact,  
the source has not gone into the quiescent state, since its discovery.
The source was active over 11 years before going to quiescence state in November, 2016 (Shaw et al. 2016; 
Plotkin et al. 2016) for $\sim 3$~months, before showing new flaring activity in February, 2017 (Kong et al. 2017; 
Qasim et al. 2017).

Recent studies of different transient BHCs show that outbursts are of two types: type-I or classical type, where
all spectral states (HS $\rightleftharpoons$ SS via two intermediate states, HIMS and SIMS) are observed, and type-II 
or harder type, where SS (sometimes even SIMS) are absent. The later type of outbursts are generally termed as `failed' outbursts.
Our recent study of few type-II outbursts of few transient BHCs, such as 2010 outburst of MAXI~J1659-152 (Debnath et al. 2015b),
2011 outburst of MAXI~J1836-194 (Jana et al. 2016), 2000 outburst of XTE~J1118+480 (Chatterjee et al. 2017) confirms 
that short orbital period BHCs generally show these class of outbursts.

The $Paper$ is organized in following way. In \S 2, we briefly discuss observation and data analysis procedure.
In \S 3, we present TCAF model fitted spectral analysis result. There we also present timing analysis results.
We also estimate mass of the BHC Swift~J1753.5-0127 based on TCAF model fitted constant normalization method. 
Finally in \S 4, a brief discussions and concluding remarks are presented.

\section{Observation and Data Analysis}

RXTE/PCA monitored the source two days after its discovery on a daily basis (Morgan et al. 2005)
Here we study archival data of $44$ observational IDs starting from the first PCA observed day; 2005 July 2 
(Modified Julian Day, i.e., MJD=53553.06) to 2005 October 19 (MJD=53662.77) using XSPEC version 12.8. 
To analyze the data we follow standard data analysis technique of the RXTE/PCA instrument 
as presented in Debnath et al. (2013, 2015a). 

For timing analysis we use the PCA {\it Science Binned} mode (FS3f*.gz) data with a maximum timing
resolution of $125\mu s$ to generate light curves for well-calibrated Proportional Counter Unit 2 
(PCU2; including all six layers) in $2-25$~keV (0-59 channels) and $2-15$~keV (0-35 channels). 
To generate the power-density spectra (PDS), ``powspec" routine of the XRONOS package is used to 
compute rms fractional variability on $2-15$~keV light curves of $0.01$~sec time bin. 
To find centroid frequencies of the observed QPOs ($\nu_{QPO}^{Obs}$) in PDS, we fit each QPO with 
a Lorentzian profile  and use the ``fit err" command to get `$\pm$' error limits.

For spectral analysis we use {\it Standard2} mode Science Data (FS4a*.gz) of the PCA instrument. 
The $2.5-25$ keV background subtracted PCU2 spectra are fitted with both the TCAF-based model {\it fits} 
file and combined disk black body (DBB) and PL model components in XSPEC. 
However, the spectra of the last eight observations, were fitted only with PL, as no significant DBB component was needed.
To achieve the best spectral fits, a Gaussian line of peak energy around $6.5$~keV (iron-line emission) is used.
Hydrogen column density (N$_{H}$) was kept frozen at 1.0$\times$10$^{21}$~atoms~cm$^{-2}$ (Kennea et al., 2011) 
for absorption model {\it wabs}. For the entire outburst, we also use a fixed $1.0$\% systematic instrumental error 
for the spectral analysis. The XSPEC command `err' is used to find 90\% confidence `$\pm$' error values for the 
model fitted parameters after achieving the best fit based on reduced chi-square value ($\chi^2_{red} \sim 1$).

To fit spectra using the TCAF-based model additive table {\it fits} file, one needs to supply five model 
input parameters such as, $i)$ black hole mass ($M_{BH}$) in solar mass ($M_\odot$) unit, $ii)$ Keplerian accretion rate 
($\dot{m_d}$ in Eddington rate $\dot{M}_{Edd}=L_E/c^2$, where $L_E$ is the Eddington luminosity and $c$ is the velocity of light),  
$ii)$ sub-Keplerian accretion rate ($\dot{m_h}$ in $\dot{M}_{Edd}$), 
$iv)$ location of the shock ($X_s$ in Schwarzschild radius $r_g$=$2GM_{BH}/c^2$) and the $v$) compression ratio. 
(R=$\rho_+ / \rho_-$, where $\rho_+$ and $\rho_-$ are post- and pre-shock densities respectively) of the shock. 
The normalization (N), though simultaneously determined is a constant and once it is determined for one outburst, 
it remains the same, though it is found to fluctuate within a narrow region due to errors in fits and
possible inaccuracies in data. Thus, once the mass ($M_{BH}$) and the normalization (N) are known, only four parameters 
are needed to fit the data unless there are precession or jet activity. The size of the hot Compton cloud, i.e., CENBOL 
and its nature (such as, height, temperature, optical depth, etc.) could be calculated from the shock parameters.

\section{Results}

Recent studies by our group show that intricacies of accretion dynamics around BHC become transparent when the data is 
analyzed with the TCAF solution. This motivated us to study the accretion flow properties of the 2005 outburst of the 
well known BHC Swift~J1753.5-0127 under TCAF paradigm. Here, we study $44$ observations between 2005 July 2 
(MJD=53553.06) and 2005 October 19 (MJD=53662.77) RXTE/PCA data for our analysis. We use RXTE/PCU2 
data for our spectral and timing analysis. To extract physical accretion flow parameters during the outburst of 
the source, $2.5-25$~keV spectra are fitted with the current version (v0.3) of the TCAF model {\it fits} file. 
We also compare our results with the spectral analysis result using DBB plus PL model. Combined DBB plus PL model 
provides only gross properties of accretion disk such as disk temperature, photon index and flux from thermal and 
non-thermal components, but real physical reasons behind the observed spectra come by fitting with TCAF solution.
From the spectral fits, we estimate the mass of the BHC, since it is also an input parameter of the present 
TCAF model {\it fits} file. 

\subsection{Spectral Analysis with the TCAF Solution and with Combined DBB and PL models}

We use $2.5-25$~keV RXTE/PCU2 data for spectral analysis of BHC Swift~J1753.5-0127 during its 2005 outburst.
Disk black body temperature ($T_{in}$) and photon index ($\Gamma$) are obtained from the combined DBB plus PL model fit 
of the spectra. We also calculated individual model component fluxes (DBB and PL separately) in $2.5-25$~keV energy band 
to have a rough estimate about the photon contribution from the thermal (DBB flux) and non-thermal (PL flux) processes around 
the BH during the 2005 outburst. Spectral fits with the current version of the TCAF solution {\it fits} 
file provide us with two component (Keplerian disk rate $\dot{m_d}$ and sub-Keplerian halo rate $\dot{m_h}$) 
accretion rates and shock parameters (location $X_s$ and compression ratio $R$).
These parameters, in turn, provide us with other physical quantities such as the viscosity, temperature,  size  
and the optical depth of the Compton cloud.

In Fig. 1a, we show $2-25$~keV PCU2 count rate variation with day (MJD), which seems to be a sharp rise and exponential 
decay type. In Figs. 1b \& 1d, variation of TCAF model fitted two component accretion rates (Keplerian disk rate 
$\dot{m_d}$, and sub-Keplerian halo rate $\dot{m_h}$) are shown. To compare in Fig. 1c \& 1e, variation of individual 
model components of the combined DBB and PL model fitted spectra are shown. 

In Figs. 2a \& 2b, we show variations of the combined DBB plus PL model fitted DBB temperature ($T_{in}$ in keV) and PL 
photon index ($\Gamma$) respectively. TCAF model fitted ratio between the sub-Keplerian halo rate and Keplerian disk rate, 
i.e., accretion rate ratio (ARR = $\dot{m_h} / \dot{m_d}$) are shown in Fig. 2c. In Fig. 2d, we show the evolution of the 
observed (primary dominating) QPO frequencies with day. 
In Fig. 3(a-d), we show variations of the TCAF model fitted shock location ($X_s$), compression ratio (R), 
normalization (N), and the derived mass.

Depending on the nature of the variations of ARR, and $\nu_{QPO}^{Obs}$, we classify the entire outburst into 
two harder spectral states in both the rising (Ris.) and declining (Dec.) phases, although many previous authors 
reported the entire outburst phase as the low HS (see, Cadolle Bel et al. 2007; Ramadevi \& Seetha 2007). The spectral 
states are also observed in the sequence: HS (Ris.) $\rightarrow$ HIMS (Ris.) $\rightarrow$ HIMS (Dec.) $\rightarrow$ HS (Dec.). 
In Fig. 4(a-d), as examples of our fits, TCAF fitted spectra with residuals for 4 observations selected from four different 
states are shown. The observation Ids are: 91094-01-01-00 (MJD=53553.05), 91094-01-01-03 (MJD=53556.19), 91423-01-02-06 
(MJD=53564.91), and 91423-01-09-00 (MJD=53612.75) are from HS (Ris.),  HIMS (Ris.), HIMS (Dec.), and HS (Dec.) respectively.
In Fig. 5(a-d), we show unabsorbed theoretical TCAF model spectra (online solid black curves) with its two components i.e., 
blackbody (online dashed blue curves) and Comptonized (online dot-dashed red curves), which are used to fit four different 
state spectra shown in Fig. 4. Photons in the multi-color blackbody type spectral component are mainly originated 
from the Keplerian disk, where as photons in the Comptonized part originate from the CENBOL. Since the Keplerian disk 
in the pre-shock region is also heated up by absorbing hard radiations from the CENBOL, it is not exactly 
the standard Keplerian disk as explained in CT95. 
Similarly, the so-called `power-law' component is not really a power-law, since it has cut-off due to recoil effect 
and a possible component reflected from the CENBOL to the observer (built-it in TCAF procedure). The number of photons 
under the PL component is those from the pre-shock Keplerian disk intercepted by the CENBOL. Thus, in effect, 
both the components are inter-linked.

\subsection{Evolution of Timings Properties during the Outburst}

Evolution of temporal properties of this BHC has been discussed by several authors (Zhang et al. 2007; Ramadevi \& Seetha 2007). 
From the outbursting profile as shown in Fig. 1a, we infer that this outburst is a Fast Rise and Slow Decay (FRSD) type rather 
than a slow rise slow Decay (SRSD; see Debnath et al. 2010).
Low frequency QPOs are observed almost every day of the 2005 outburst of Swift~J1753.5-0127. In the rising phase of the outburst, 
QPOs are observed on all $5$ observations, where as in the declining phase, QPOs are observed in 28 observations out of total $39$ 
observations. Most of the observed QPOs are strong type-C QPOs. 

A sharp rise in X-ray intensity (see, Fig. 1a) and QPO frequencies (from $0.622$ to $0.813$ Hz) are observed in the initial two days, 
from 2005 July 2 (MJD=53553.06) to 2005 July 4 (MJD=53555.21), then slow rise until 5th observation i.e., 2005 July 6 (MJD=53557.24) 
of the outburst. After that, both PCU2 count rates and QPO frequencies are showed a general trend of slow decreasing nature 
till the end of our observations. 2005 September 17 (MJD=53630.31) is the last day during the outburst, where a prominent QPO of 
$0.193$~Hz is observed.

\subsection{Evolution of Spectral Properties during the Outburst}

We use $2.5-25$ keV RXTE/PCU2 data for the spectral analysis. With TCAF solution, we extract physical parameters 
such as two types of accretion rates ($\dot{m_d}$ and $\dot{m_h}$) in Eddington accretion rate unit, 
shock location ($X_s$) in ($r_g$) unit and compression ratio ($R$) from each spectral fit. From combined DBB and PL model 
fitted spectra, we obtain the value of PL photon index ($\Gamma$), DBB temperatures ($T_{in}$ in keV). Based on the 
variations of the ARRs (=$\dot{m_h}/\dot{m_d}$) and nature (shape, frequency, $Q$ value, rms\% etc) of QPOs, 
we find only two harder (HS and HIMS) spectral states during the 2005 outburst of the BHC Swift~J1753.5-0127. 
Similar to 2011 outburst of MAXI~J1836-194 (Jana et al. 2016), these spectral states are observed in the 
sequence of : HS (Ris.) $\rightarrow$ HIMS (Ris.) $\rightarrow$ HIMS (Dec.) $\rightarrow$ HS (Dec.). 
The nature of the source during observed spectral states are discussed below.

\subsubsection{Hard State - Rising}

Initial two observations (from 2005 July 2 to 4; MJD=53553.06 to 53555.21) belong to this spectral state. During this phase 
of $\sim 2$~days, a sharp rise in the TCAF model fitted both accretion rates ($\dot{m_d}$ and $\dot{m_h}$) as well as individual 
model flux components of the combined DBB and PL model fits are observed. QPO frequency also rises sharply during these two 
observations. Similar to other transient BHCs (e.g., H~1743-322, GX~339-4, MAXI~J1659-152, MAXI~J1836-194), recently 
studied by our group with the TCAF solution (see, Mondal et al. 2014; Debnath et al. 2015a,b; Jana et al. 2016), a local 
maximum in ARR values is also observed on the 2nd observation of the current outburst of Swift~J1753.5-0127. 
So, we define this observation as the transition from HS (Ris.) to HIMS (Ris.).

\subsubsection{Hard-Intermediate State - Rising}

The source has been observed to this spectral state for the next two days (from 2005 July 4 to 6; MJD=53555.21 to 53557.24), three 
observations. During this phase of the outburst, a slow rise in both disk and halo accretion rates are observed, although rise in the 
disk rate is slightly higher than halo rate. As a result of the ARR values are observed to decrease slowly, and becomes lowest on the 
3rd observation day (MJD=53557.24). We define this day as the transition day from HIMS (Ris.) to HIMS (Dec.). After that, X-ray 
intensities as well as both the components of accretion rates start to decrease. 
Also, on this day, maximum values of the PL photon 
indices ($\Gamma$=$1.71$) and QPO frequencies ($\nu_{QPO}^{Obs}$=$0.888$~Hz) during entire outburst are observed. 

\subsubsection{Hard-Intermediate State - Declining}

The source was observed in this spectral state for around next $\sim 32$~days. During this phase of the outburst, X-ray intensity 
as well as TCAF model fitted two component accretion rates decrease monotonically, although there is a rise in ARR values. 
This indicates that the rate of fall in Keplerian disk rate is faster than the sub-Keplerian halo rates, which implies that 
the spectrum becomes harder as day progresses. This hardening of the spectra also could be concluded from the variation of 
the PL indices. A general decreasing nature of the $\Gamma$ values are observed during this phase of the outburst. 

The shock is found to move slowly away from the BH, while compression ratio (R) stays roughly at a constant value of 
$\sim 1.07-1.22$. QPOs are also observed to decrease with time (day). We define observation on the 2005 August 7 (MJD=53589.49) 
as the transition day from HIMS (Dec.) to HS (Dec.). On this particular observation, the shock becomes stronger with $R=1.745$. 
After that, ARRs, QPO frequencies, PL indices became roughly constant and the trend continued till the end of our observation. The shock 
started to recede back with an acceleration with progressively increasing shock strength after the transition day.

\subsubsection{Hard State - Declining}

The source is found in this spectral state till the end of our observation starting from the HIMS (Dec.) to HS (Dec.) 
transition day (MJD=53589.49). During this phase of the outburst, both TCAF model fitted accretion rates decrease roughly 
in the same rate due to lack of supply of matter from the companion. As a result of that, ARRs are observed roughly at 
a constant value of $\sim 0.52$. QPOs are also observed roughly at a constant frequency (within a short range of 
$\sim 0.19-0.23$~Hz), although they have not been detected for all observations during this phase of the outburst.
During the initial period of this phase, the shock moves away from the BH with increasing shock strength, before it 
settles down into a roughly constant value starting from 2005 August 25 (MJD=53607.25). The outburst continued for 
a long duration even after our observations with low X-ray activity. 

\subsection{Estimation of the Mass of the Black Hole}

Recently, MDCMJ16 used two independent methods, such as, $i)$ TCAF model fitted constant normalization (N), and 
$ii)$ propagating oscillatory model (POS) model (see, Chakrabarti et al. 2008; Debnath et al. 2010, 2013; Nandi et al. 2012) 
fitted QPO frequency evolution, to measure unknown mass of BHC MAXI~J1659-152. Furthermore, these methods are successfully 
applied to measure mass of other BHCs of unknown mass, such as, MAXI~J1836-194 (JDCMM16), MAXI~J1543-564 (CDCMJ16), 
H~1743-322 (Bhattacharjee et al. 2017; Molla et al. 2017), which motivated us to measure the mass of the current 
BHC Swift~J1753.5-0127 using constant normalization method. 

We believe that TCAF model fitted N values should be constant for a particular source, since it only depends on physical 
parameters such as the mass, distance, disk inclination angle, etc. Here, during the entire 2005 outburst of Swift~J1753.5-0127, 
TCAF normalization show a constant signature of $\sim 1.6$, except for the high values ($\ge 2.0$) for $5$ observations 
in the initial period of HIMS (Dec.). A significant deviation of the N value could be due to high X-ray emissions from 
the jet (Soleri et al. 2010), whose effects have not been included in the current version of the TCAF model fits file. 
Emission from the jet will be discussed in another work (Jana et al. 2017). 
Mass of the BH ($M_{BH}$) is an important parameter of the TCAF model {\it fits} file, which was kept free for all 
observations. From the spectral fit with the TCAF solution, we find that $M_{BH}$ varies in between 
$4.75 M_\odot$ to $5.90 M_\odot$. This is our estimated range of the probable mass the source. 

\section{Discussions and Concluding Remarks}

We study spectral and temporal evolutions of BHC Swift~J1753.5-0127 during its 2005 outburst to infer 
about the accretion flow properties of the source. For temporal study, we use $2-25$~keV RXTE/PCU2 data and 
for spectral analysis $2.5-25$~keV data of the same instrument for $44$ observation Ids are used. 
Low frequency QPOs and their nature, such as the shape, frequency, $Q$ value, rms\%, etc. are studied during the
entire 2005 outburst of the BHC Swift~J1753.5-0127. 
To study spectral properties, spectra are fitted with the current version (v0.3) of the TCAF solution {\it fits} 
file to extract physical accretion flow parameters, such as, Keplerian disk rate ($\dot{m_d}$), sub-Keplerian 
halo rate ($\dot{m_h}$) in Eddington rate, shock location ($X_s$) in $r_g$ and shock compression ratio ($R$), 
and mass of the BH. A comparison is made with the spectral analysis result done with the combined DBB and PL models.
A combined analysis of DBB and PL model gives us a rough estimation of disk temperature, photon indices and 
thermal and non-thermal flux contributions, etc. However, a TCAF model fits provide us 
with the evolution of the physical parameters relating to the accretion flow itself. 
Variation of TCAF extracted parameters together with ARR, and observed QPOs help us to determine 
the spectral state of the BHC. We conclude that this source has been in two harder (HS and HIMS) spectral states 
during the 2005 outburst, although there are many reports from other authors that during the entire outburst phase 
source was in low HS (see, Cadolle Bel et al. 2007; Ramadevi \& Seetha 2007). 
No signature of softer (SIMS and SS) spectral states are observed by us. The reason may be the short orbital period 
($3.2$~hrs, Zurita et al. 2007) of the binary system which makes the system very compact. There is a general 
excess of matter surrounding the whole system. So, there may be a constant supply of sub-Keplerian matter from 
the wind. This is also the reason that the outburst continued till date in HS with low X-ray intensity.
Since during the outburst, this source does not follow canonical hysteresis loop of HID 
or `q'-diagram or ARRID (see, JDCMM16 and references therein), as it never visited 
softer spectral states, the 2005 outburst of Swift~J1753.5-0127 may be termed as 
a `failed' outburst. Several other BHCs, such as, MAXI~J1659-152, MAXI~J1836-194, 
XTE~J1118+480, Swift~J1357.2-0933 also have shown similar behavior during their outbursts.

Similar to MAXI~J1836-194 during its 2011 outburst (JDCMM16), the spectral states are observed in the sequence of: 
HS (Ris.) $\rightarrow$ HIMS (Ris.) $\rightarrow$ HIMS (Dec.) $\rightarrow$ HS (Dec.) to form a hysteresis loop without 
softer spectral states. X-ray intensities, QPO frequencies as well as two component accretion rates, PL indices rise 
during initial five days of the outburst. This phase of the outburst is divided into 
HS (Ris.) and HIMS (Ris.) with a transition on the 2nd observation (2005 July 4; MJD=53557.24), when a sharp rise in above mentioned 
parameters from the first observation and a local maximum of the ARR are observed. This property is found to be a common feature on 
HS $\rightleftharpoons$ HIMS transition days, observed for other transient BHCs, such as H~1743-322, GX~339-4, MAXI~J1659-152, 
MAXI~J1836-194 during their outbursts (see, MDC14, DMC15, DMCM15, JDCMM16).
After that, X-ray intensity, QPO frequency as well as two component accretion rates are observed to decrease with time (day) till 
the end of our observations. We define this phase of the outburst into two declining harder spectral states, i.e., HIMS (Dec.) and HS (Dec.). 
The transition between these spectral states is defined on 2005 August 7 (MJD=53589.49), since after that QPO frequencies, ARRs, PL indices 
become roughly constant. Our definition of 2005 July 6 (MJD=53557.24) observation as the transition between two hard-intermediate 
spectral states is clearly justified as on this particular observation maximum values of X-ray intensities, QPO frequencies, 
Keplerian rates are observed. 

QPO frequencies monotonically increase  during the rising phase of the outburst, whereas during the declining phase of the outburst 
a general trend of decreasing nature of QPO frequencies is observed though on a daily basis the decrease is not monotonic. There 
may be some turbulence present in the flow. According to the TCAF solution, origin of these low frequency QPOs could easily be 
explained by the resonance oscillation (type-C), weak oscillation (type-B) of the Compton cloud boundary (i.e., shock) or even 
of the shock-free centrifugal barrier (type-A). So, we suspect that during the non-QPO days (in the declining HS), resonance 
condition between cooling and infall time of the post-shock region was not satisfied by a large margin 
(see, Chakrabarti et al., 2015 for more details).

Through TCAF fits, we find that an outburst of a transient BHC is manifested due to a large supply of Keplerian and sub-Keplerian
matter close to the inner disk. It is reasonable to assume that this is due to rapid rise in viscosity, perhaps of magnetic origin, 
owing to enhanced magnetic activity in the companion or convective turbulence at the outer disk (Chakrabarti 2013). Similarly, 
declining phase of an outburst starts when supply from the companion is turned off, i.e., due to low supply of the high viscous matter. 
Although the 2005 outburst of Galactic BHC Swift~J1753.5-0127 started in the canonical way, it did not move to the burst-off 
phase for over a decade now. This may be due to the constant supply of matter from the companion star, most probably in the form 
of wind flow of low angular momentum low viscous sub-Keplerian halo matter. 
For this reason, sometimes even for high accretion rate in the Keplerian component could not able to cross the comparative 
limiting value, such that source reaches soft spectral state.
The binary system, Swift~J1753.5-0127 seems to belong to a type-II class of outbursting sources along with MAXI~J1836-194, 
MAXI~J1659-152, XTE~J1118+480, etc. of shorter orbital periods, where softer spectral states are also missing during their outbursts. 

For comparison of parameters obtained from TCAF fits with those we have obtained 
from DBB \& PL model, we compute the $R_{in}$, the inner edge of the so-called truncated standard disk. 
In TCAF fits, $X_s$ is the shock location which is the inner boundary of a modified standard disk
and the outer boundary of the CENBOL. $R_{in}$ (in $km$) computed from DBB normalization from the relation  
$N_{DBB}=(R_{in}^{2}/ D^{2})cos (i)$, where, $D$ is in $10$ $kpc$ and $i$ is inclination angle,
is  given in the Table in Appendix -I. The normalization is used for
the whole spectrum to match photons emitted at the disk frame and those observed by us.
$R_{in}$, calculated from DBB norm, is not really a physical parameter. 
For instance, we see from the Appendix Table that on MJD=53583, $R_{in}$ is less than 30 km. 
For a BH mass of $5.5$ $M_{\odot}$, $R_{in}$ goes inside the 
event horizon which is not possible. Similar value of $R_{in} < 2 r_{g}$ was found in 
seven more observations. In contrast, normalization in TCAF is constant
across the spectral states and the size of the Compton cloud $X_s$ could be 
tens to hundreds of Schwarzschild radii away. In other word, TCAF solves for totally different
physical parameters of the system than those from the DBB \& PL model.

It is important to note that according to TCAF, all the four parameters (other than the mass and 
the Normalization which do not change in principle on a daily basis, unless there are addition sources of photons 
from jets), are essential to fit any general data from a stellar mass black hole having no outflows or magnetic 
field. In certain period of time, one or more parameters may be constant or nearly constant as special case. 
For instance in soft/very soft states, only accretion rate of the disk ${\dot m}_d$ would be required, and two  
components are not necessary. Fitting with DBB plus PL models often requires an additional `reflection' component 
to take care of a `bump' like features. In TCAF this is automatically produced when the Keplerian disk 
is heated up by the fraction of incident Comptonized photons from the CENBOL. Thus we do not require this 
additional component. It is not unlikely that this reprocessing may not fully account for the reflection component, 
especially when the disk is partly ionized. However, in our case the Keplerian disk is flanked by a hot advective flow, 
and thus this problem of partial ionization, especially closer to the black hole, does not arise.

All the four parameters play crucial role in fitting a spectrum even it appears to have simple features. 
The coupled differential equations given in CT95 produce a spectrum in a self-consistent way. 
Keplerian disk component, depending on its accretion rate ${\dot m}_d$, emits seed photons, a fraction 
of which are intercepted by the CENBOL. The fraction depends on the optical depth, height (related to 
the CENBOL temperature) of the CENBOL, and hence depends on all four parameters. The degree of Comptonization 
of the seed photons will decide the PL slope of the spectrum. If the CENBOL size ($X_s$) gets reduced, it it 
expected that it should be easier to cool it since the optical depth is much higher in a denser region. 
However, the reduction in size reduces the path length as well as the number of intercepted seed photons 
from the Keplerian flow. The Compression ratio ($R$) increases the density in CENBOL by this  factor, 
causing more Comptonization. The internal density of the CENBOL depends on the sum of the two rates 
(${\dot m}_d$ and ${\dot m}_h$) of the accretion flow. As the CENBOL cools down with higher ${\dot m}_d$, 
the possibility of thermal pressure driven jets from CENBOL becomes lesser. Thus in TCAF scenario, Jets 
and outflows are possible in harder states.

In sort, all the four parameters play their roles in a complex and non-linear way. It is true that 
when the disk rate is very small, as shown in CT95, the spectral slope is fairly insensitive to 
the rate, till the time the disk rate becomes comparable to the halo rate to cool CENBOL down. So one 
would imagine that the ${\dot m}_d$ would be degenerate. This is not true, since the same rate also 
is responsible for the multi-color blackbody bump. On the other hand, the resulting spectral slope 
would be different when the mass changes, since  changes the seed photon temperature, without changing 
the advective flow properties. Hence the information of mass enters through the spectral properties of the 
softer component. Since TCAF obtains the entire spectrum at a time at the rest frame of the emitter, only 
one normalization is needed to map the whole spectrum with the observed spectrum. There could be fluctuations 
due to observational errors, but our model normalization is not expected to change violently for the same object 
unless there are addition source of X-rays from the base of the outflows and jets. 
Thus the average normalization obtained from one outburst is expected to fit the data in other outbursts as well.

Using TCAF, we also estimated probable range of the mass of the BHC from our TCAF model fitted spectral study. Mass of the BH 
is an important parameter of the current TCAF model {\it fits} file, which was kept free while fitting spectra. We assumed the 
model normalization to remain constant (as in MDCMJ16) to estimate mass of the source to be about $4.75 M_{\odot}$ to $5.90 M_{\odot}$. 
TCAF model normalization (N) does not vary significantly for a particular BHC even when it passes through different outbursts 
and spectral states, unless there is a precession in the disk which may change the projected surface of the emission area or 
there are some significant outflow activities which we missed to include in current model {\it fits} file. During the entire 2005 outburst 
of Swift~J1753.5-0127, model normalization is found to be constant at $\sim 1.6$ with a little variation in between $1.41-1.81$. 
A large deviation (N $\ge 2.0$) is observed in five observations during the HIMS (Dec.), 
and this may be due to high Jet emissions (Soleri et al. 2010) during that period of observations, whose effects were not  
included in the current version of the TCAF model fits file. In future, we will calculate X-ray flux coming from the jet 
and study their variations during the outburst using the variation of N during an outburst. This will be published elsewhere.
A broad band X-ray study of the source starting from July 2005 till date, 
using archival data of RXTE (PCA and HEXTE instruments) and SWIFT 
(XRT and BAT instruments) satellites will also be done and published elsewhere.

\section*{Acknowledgments}

A.J. and D.D. acknowledge support from ISRO sponsored RESPOND project fund (ISRO/RES/2/388/2014-15). 
D.C. and D.D. acknowledge support from DST sponsored Fast-track Young Scientist project fund (SR/FTP/PS-188/2012).
SM acknowledge supports FONDECYT post-doctoral fellowship grand (\# 3160350).

{}

\clearpage


\begin{figure}
\vskip -0.5cm
\centerline{
\includegraphics[scale=0.6,angle=0,width=8.0truecm]{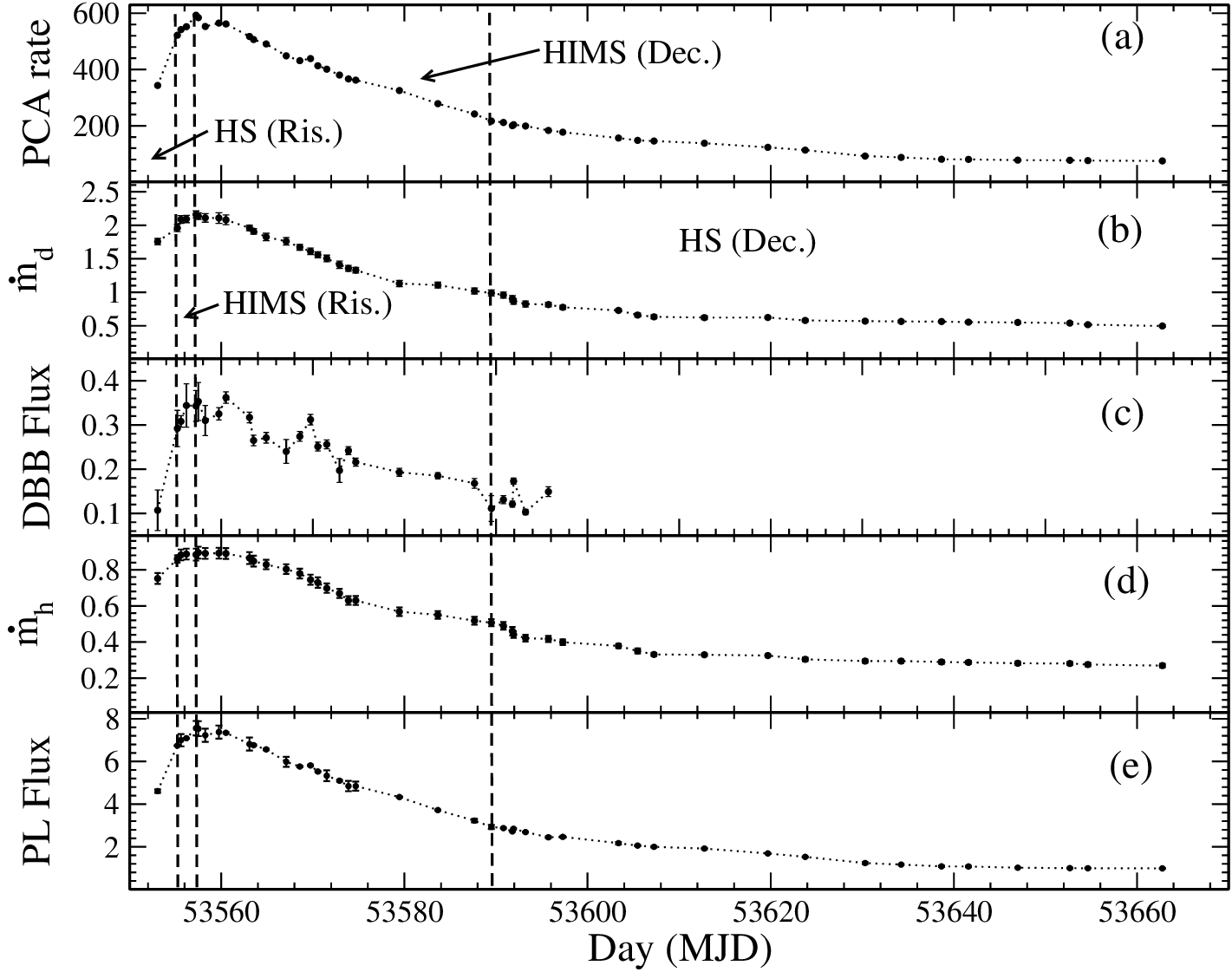}}
\caption{Variation of (a) $2-25$~keV X-ray intensity i.e., PCU2 count rate ($cnts/sec$) is shown in the top panel. 
Variations of TCAF model ($2.5-25$~keV spectral) fitted Keplerian disk rate ($\dot{m_d}$) in $\dot{M}_{Edd}$, and 
sub-Keplerian halo rate ($\dot{m_d}$) in $\dot{M}_{Edd}$ are shown in panels (b) and (d) respectively. Similarly, 
variations of DBB+PL model fitted DBB and PL flux components in the $2.5-25$~keV energy band are shown in panels 
(c) and (e) respectively.}
\label{fig1}
\end{figure}

\begin{figure}
\vskip -0.5cm
\centerline{
\includegraphics[scale=0.6,angle=0,width=8.0truecm]{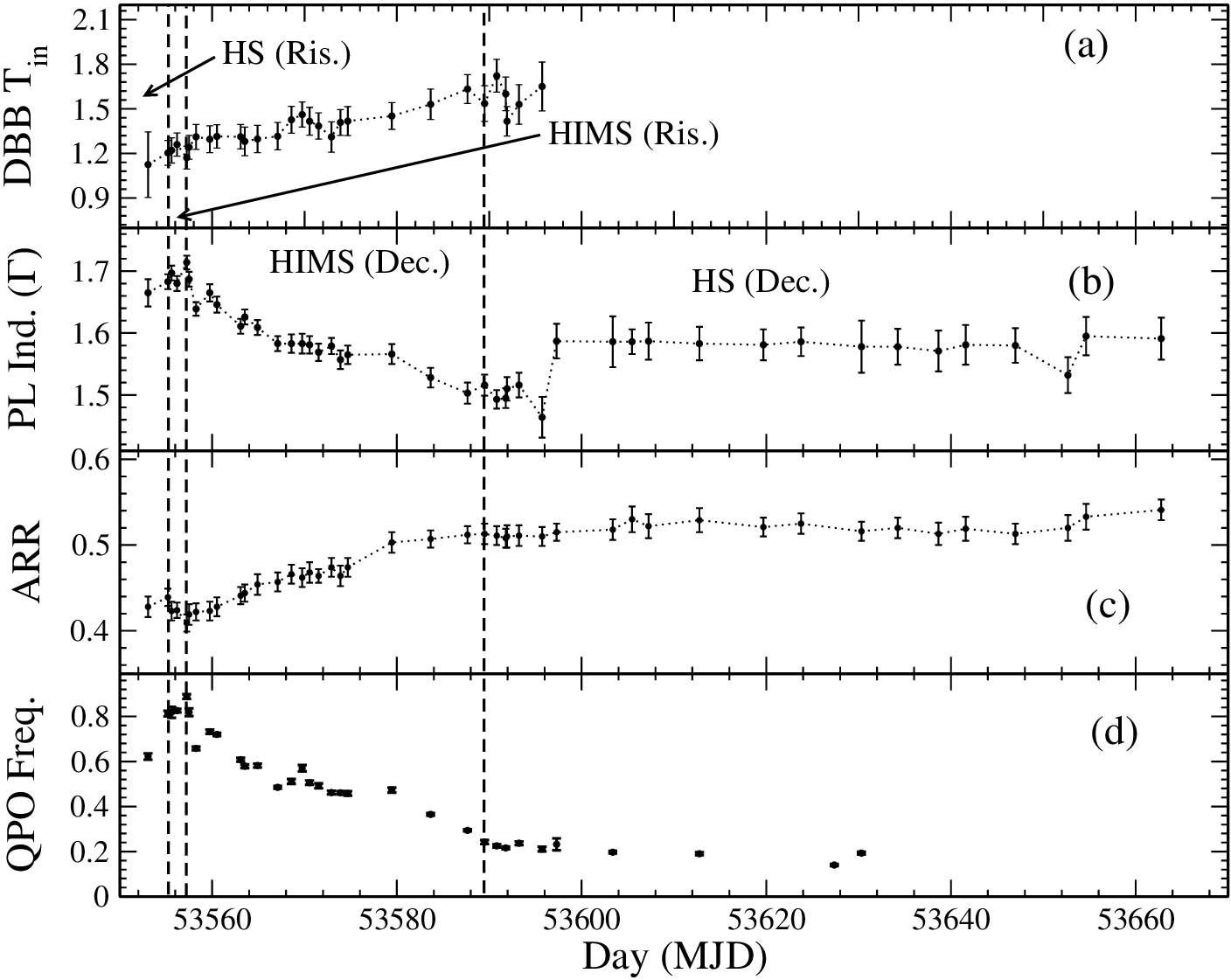}}
\caption{(a-b) Combined DBB plus PL model fitted DBB temperature ($T_{in}$ in keV) and PL photon index ($\Gamma$) are 
shown in top two panels. Variations of (c) accretion rate ratio, i.e. ARR ($=\dot{m_h} / \dot{m_d}$), and (d) observed 
QPO frequencies (in Hz) are shown in bottom to panels.}
\label{fig2}
\end{figure}

\clearpage
\begin{figure}
\vskip -0.5cm
\centerline{
\includegraphics[scale=0.6,angle=0,width=8.0truecm]{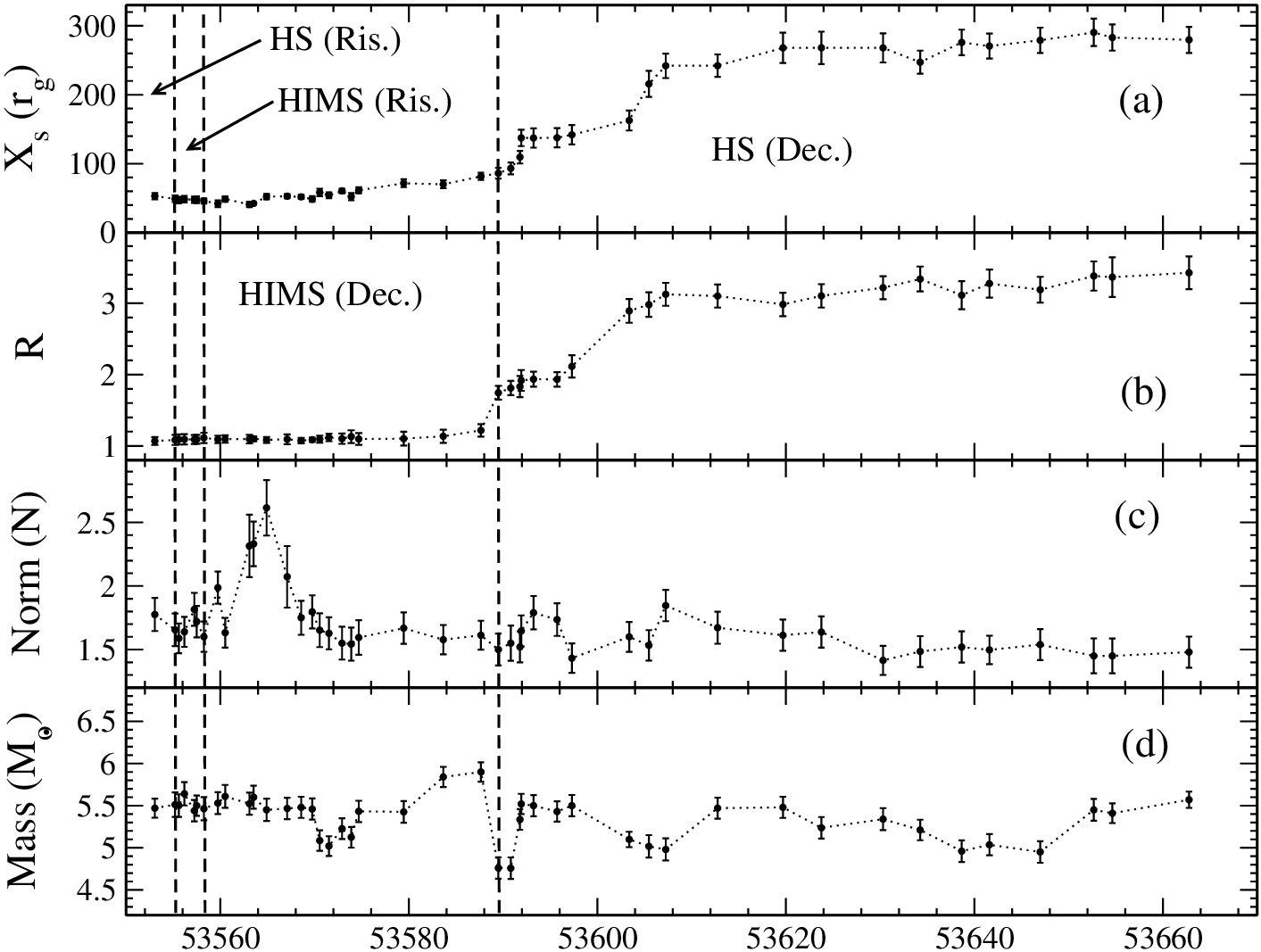}}
\caption{(a-b) Variations of TCAF model fitted shock parameters (location $X_s$ in $r_g$ and compression ratio)
are shown in the top two panels. In the bottom two panels, variations of the TCAF model fitted normalization (N) 
and BH mass (in $M_\odot$) values are shown.}
\label{fig3}
\end{figure}

\begin{figure}
\vskip -0.5cm
\centerline{
\vbox{
\includegraphics[scale=0.4,width=5.5truecm,angle=270]{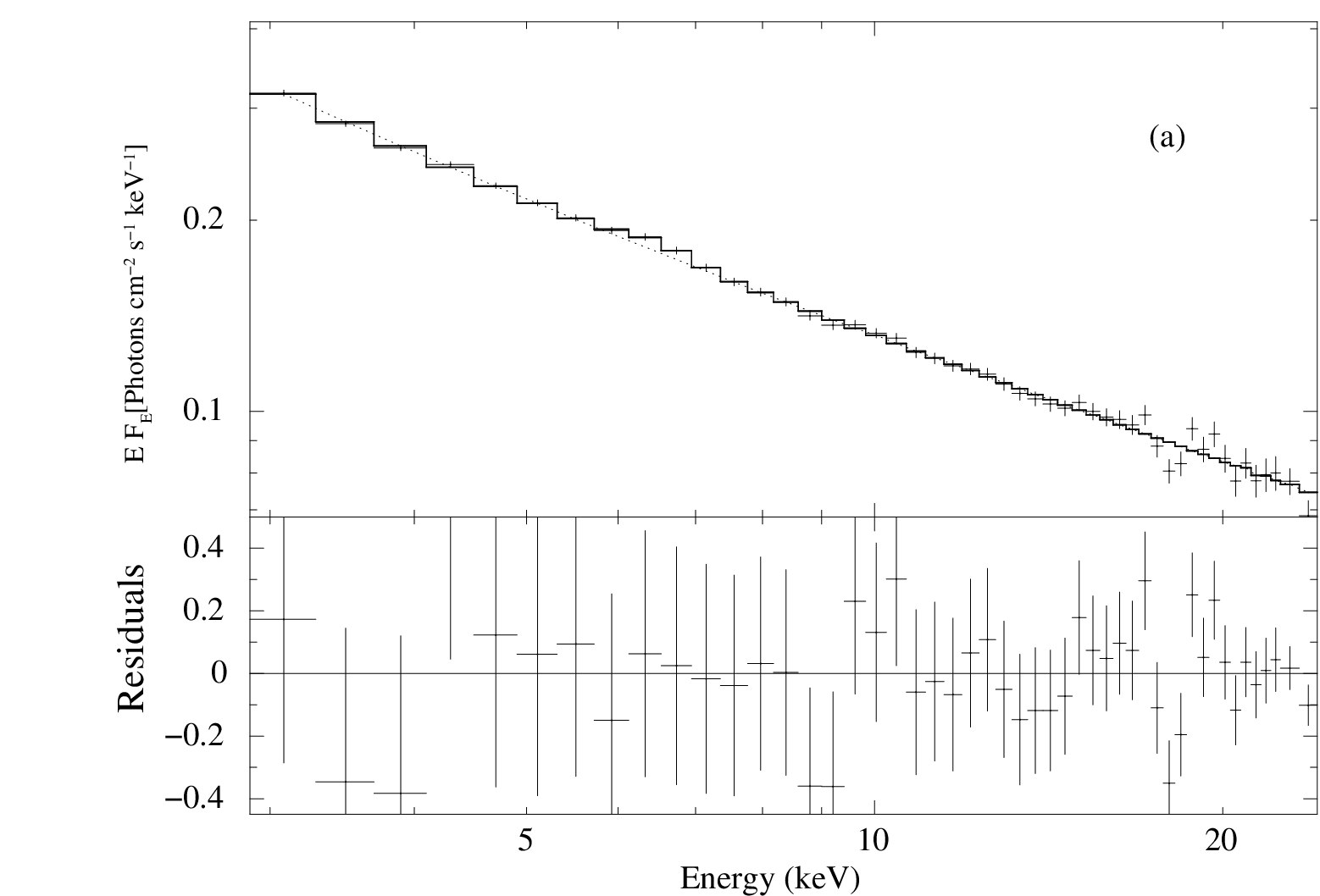} \vskip 0.05cm
\includegraphics[scale=0.4,width=5.5truecm,angle=270]{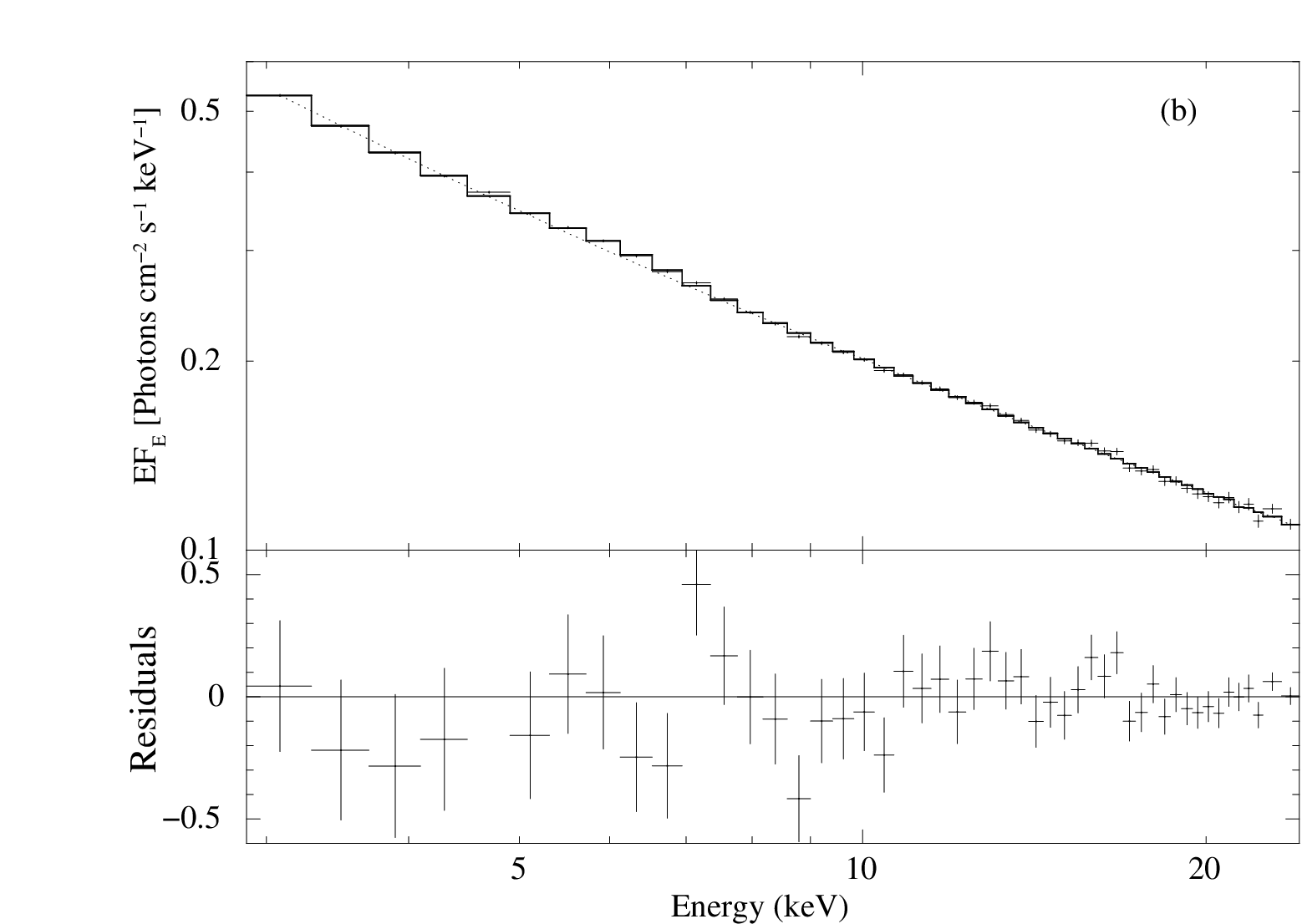} \vskip 0.05cm
\includegraphics[scale=0.4,width=5.5truecm,angle=270]{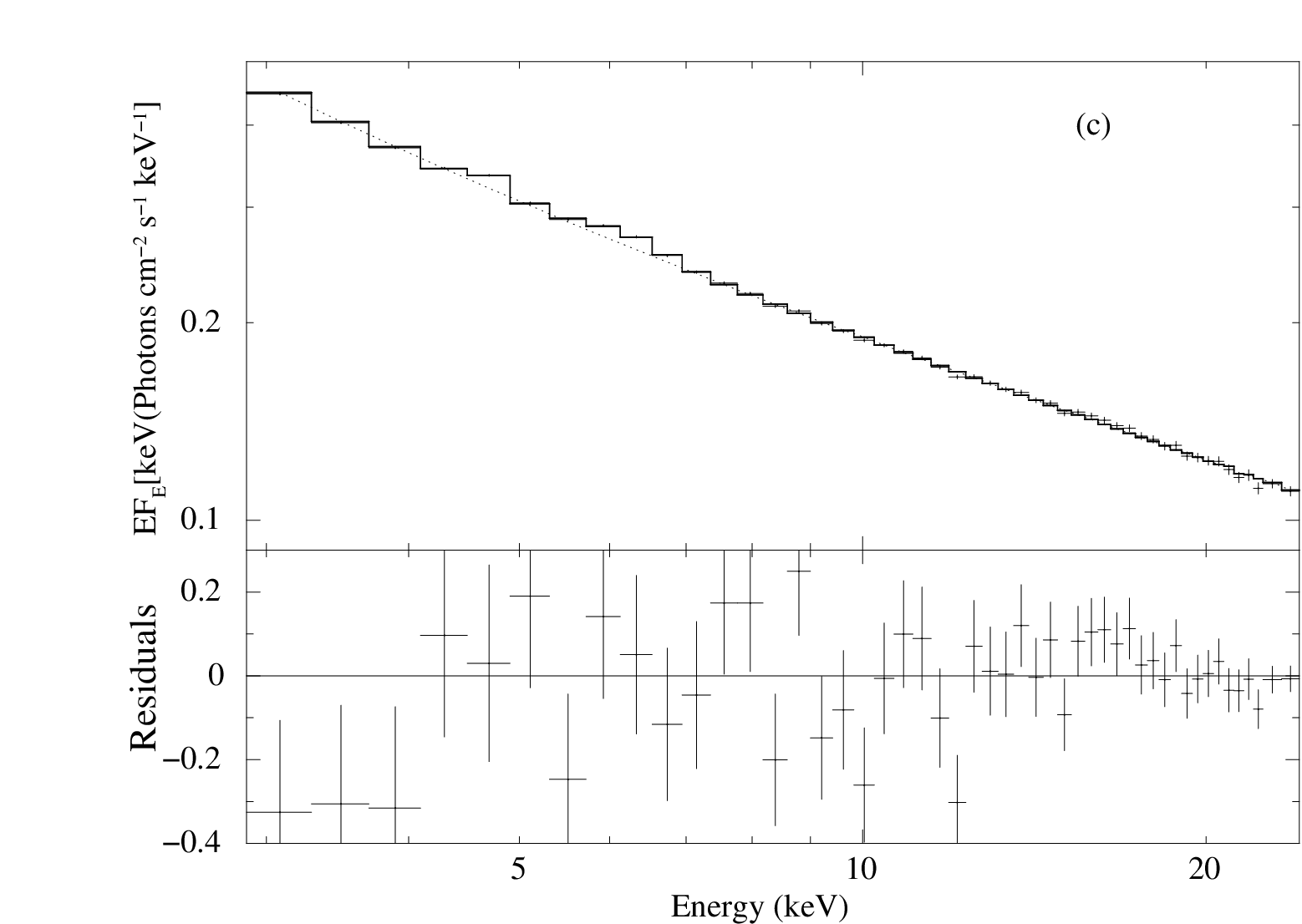} \vskip 0.05cm
\includegraphics[scale=0.4,width=5.5truecm,angle=270]{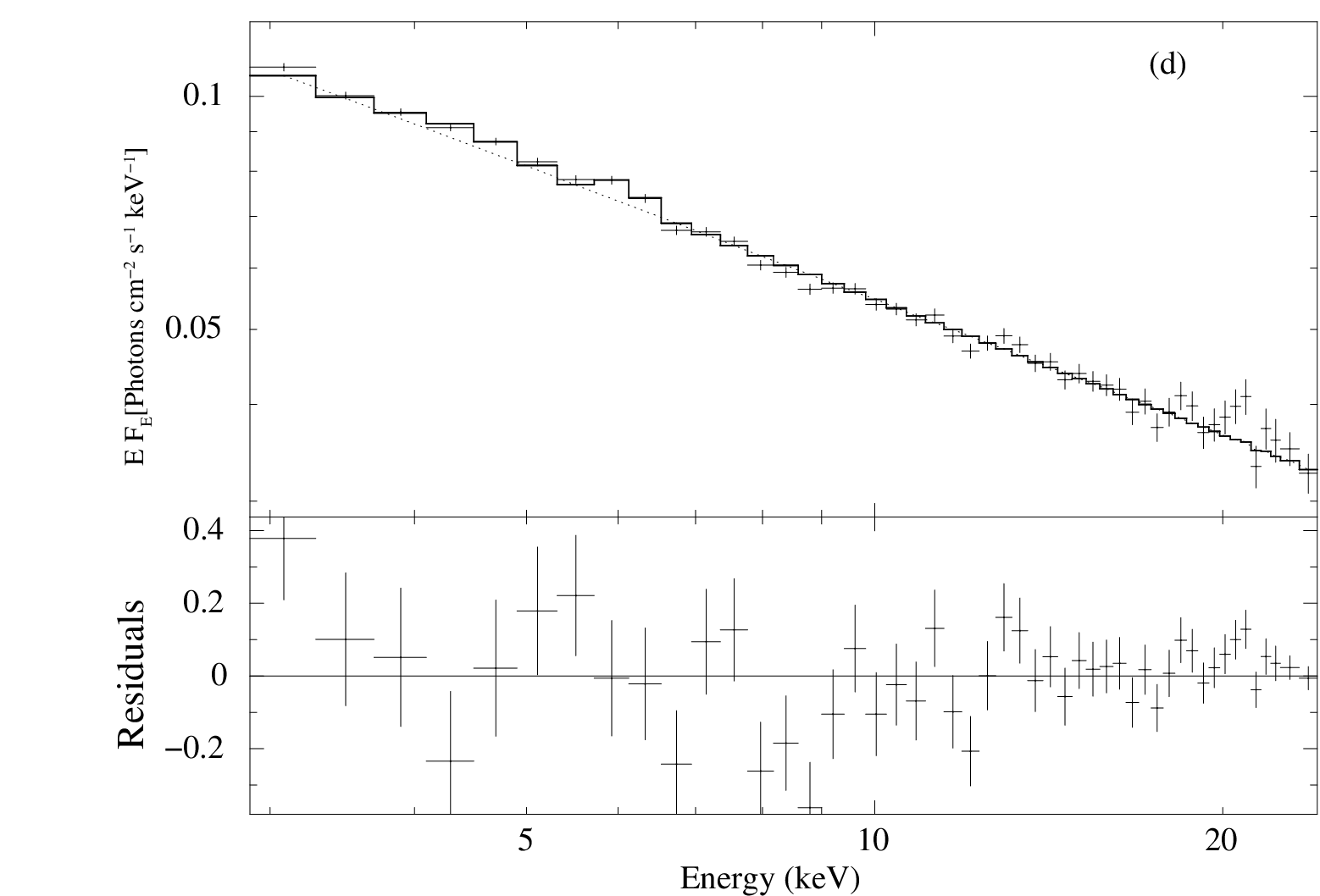} }}
\vskip 0.1cm
\caption{ TCAF model fitted spectra from four different states: HS (Ris.), HIMS (Ris.), HIMS (Dec.), and HS (Dec.) for
observation ID: (a) 91094-01-01-00, (b) 91423-01-01-03, (c) 91423-01-02-06, and (d) 91423-01-09-00 respectively.}
\label{fig4}
\end{figure}

\begin{figure}
\vskip -0.5cm
\centerline{
\includegraphics[scale=0.6,angle=0,width=8.0truecm]{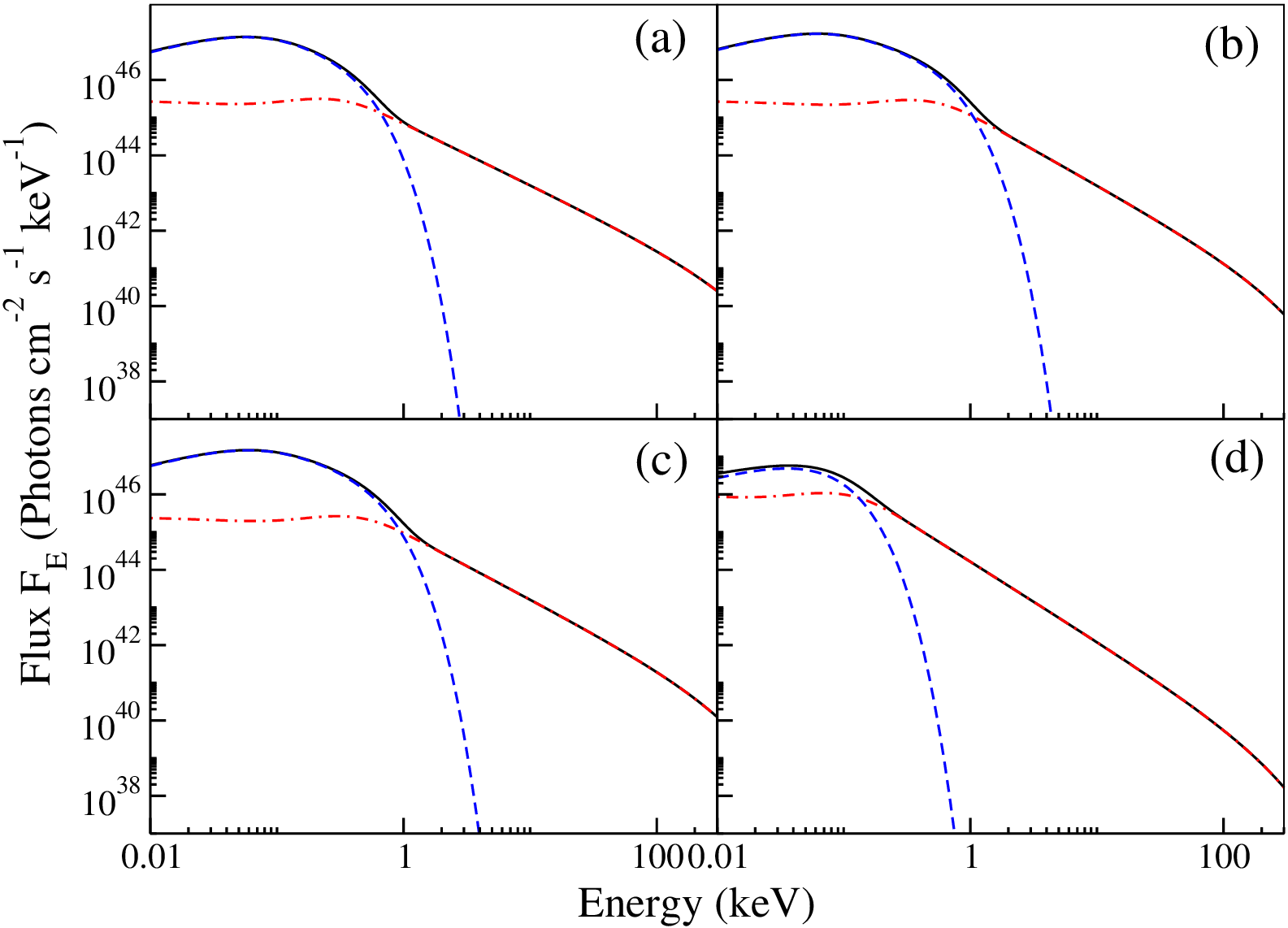}}
\caption{Blackbody (online dashed blue curves), and Comptonized components (online dot-dashed red curves) 
of the unabsorbed TCAF theoretical spectra, which are used to fit four different states spectra of Fig. 4 are shown. 
The total spectra are shown in solid black curves. Thermal blackbody photons primarily originates from the Keplerian disk 
and Comptonized photons originated from `hot Compton cloud' i.e., from CENBOL, which indeed forms due to sub-Keplerian 
halo matters.}
\label{fig5}
\end{figure}

\clearpage

\begin{table}
\vskip -2.0cm
\addtolength{\tabcolsep}{-4.50pt}
\scriptsize
\centering
\centering{\large \bf Appendix I}
\vskip 0.2cm
\centerline {2.5-25 keV Combined DBB plus PL Model and TCAF Model Fitted Spectral Parameters with QPOs}
\vskip 0.2cm
\begin{tabular}{lcccccccccccccccc}
\hline
Obs. & Id&MJD&$T_{in}$&$\Gamma$&$DBBf^\dagger$&$PLf^\dagger$&$R_{in}$&$M_{BH}$&$\dot{m_d}$&$\dot{m_h}$&ARR&$X_s$&R&Norm.&QPO$^{\dagger\dagger}$&$\chi^2/DOF$\\
   & & & (keV) & & & &(km) & $M_{\odot}$ &($\dot{M}$$_{Edd}$)&($\dot{M}$$_{Edd}$)&&($r_g$)  & &   \\
 (1)&  (2)  & (3)  & (4)& (5) & (6) & (7) &  (8) & (9) & (10) & (11) & (12) & (13)  & (14) & (15) & (16) & (17)\\
\hline
 1&X-01-00&53553.06&$1.125^{\pm 0.221}$&$1.665^{\pm 0.022}$&$0.107^{\pm 0.221}$&$4.609^{\pm 0.079}$&$42.79^{}$&$5.47^{\pm 0.31}$&$1.755^{\pm 0.048}$&$0.752^{\pm 0.030}$&$0.428^{\pm 0.029}$&$52.93^{\pm  4.71}$&$1.069^{\pm 0.055}$&$1.776^{\pm 0.130}$&$0.622^{\pm 0.014}$&31.17/41 \\   
  & & & & & & & & & & & & & & & \\
 2&X-01-01&53555.21&$1.204^{\pm 0.084}$&$1.683^{\pm 0.012}$&$0.292^{\pm 0.083}$&$6.743^{\pm 0.014}$&$59.18^{}$&$5.51^{\pm 0.25}$&$1.959^{\pm 0.055}$&$0.860^{\pm 0.020}$&$0.439^{\pm 0.022}$&$49.06^{\pm  4.75}$&$1.083^{\pm 0.071}$&$1.656^{\pm 0.128}$&$0.813^{\pm 0.012}$&32.75/41 \\
 3&X-01-02&53555.60&$1.221^{\pm 0.085}$&$1.697^{\pm 0.012}$&$0.309^{\pm 0.084}$&$6.998^{\pm 0.289}$&$58.76^{}$&$5.50^{\pm 0.34}$&$2.086^{\pm 0.058}$&$0.882^{\pm 0.030}$&$0.423^{\pm 0.026}$&$47.37^{\pm  4.75}$&$1.090^{\pm 0.071}$&$1.589^{\pm 0.117}$&$0.818^{\pm 0.025}$&22.89/41 \\ 
 4&X-01-03&53556.19&$1.260^{\pm 0.078}$&$1.680^{\pm 0.012}$&$0.344^{\pm 0.077}$&$7.089^{\pm 0.017}$&$57.17^{}$&$5.64^{\pm 0.39}$&$2.094^{\pm 0.055}$&$0.887^{\pm 0.030}$&$0.424^{\pm 0.025}$&$48.55^{\pm  4.64}$&$1.092^{\pm 0.072}$&$1.639^{\pm 0.117}$&$0.826^{\pm 0.008}$&30.17/40 \\ 
  & & & & & & & & & & & & & & & \\
 5&Y-01-04&53557.24&$1.172^{\pm 0.077}$&$1.714^{\pm 0.011}$&$0.343^{\pm 0.077}$&$7.559^{\pm 0.336}$&$68.79^{}$&$5.44^{\pm 0.28}$&$2.162^{\pm 0.052}$&$0.884^{\pm 0.033}$&$0.409^{\pm 0.025}$&$47.65^{\pm  4.63}$&$1.090^{\pm 0.063}$&$1.816^{\pm 0.129}$&$0.888^{\pm 0.010}$&23.96/41 \\ 
 6&X-01-04&53557.50&$1.242^{\pm 0.080}$&$1.687^{\pm 0.012}$&$0.353^{\pm 0.079}$&$7.537^{\pm 0.349}$&$60.11^{}$&$5.50^{\pm 0.20}$&$2.137^{\pm 0.050}$&$0.895^{\pm 0.033}$&$0.419^{\pm 0.025}$&$47.67^{\pm  4.82}$&$1.092^{\pm 0.062}$&$1.720^{\pm 0.123}$&$0.819^{\pm 0.017}$&39.80/41 \\ 
 7&Y-01-00&53558.25&$1.312^{\pm 0.084}$&$1.639^{\pm 0.011}$&$0.311^{\pm 0.084}$&$7.226^{\pm 0.310}$&$49.10^{}$&$5.46^{\pm 0.38}$&$2.110^{\pm 0.064}$&$0.891^{\pm 0.030}$&$0.422^{\pm 0.027}$&$46.04^{\pm  3.85}$&$1.112^{\pm 0.073}$&$1.661^{\pm 0.118}$&$0.658^{\pm 0.008}$&39.95/40 \\ 
 8&X-02-01&53559.74&$1.296^{\pm 0.090}$&$1.665^{\pm 0.014}$&$0.325^{\pm 0.090}$&$7.379^{\pm 0.305}$&$51.85^{}$&$5.53^{\pm 0.29}$&$2.109^{\pm 0.079}$&$0.892^{\pm 0.030}$&$0.423^{\pm 0.030}$&$42.00^{\pm  4.86}$&$1.091^{\pm 0.054}$&$1.987^{\pm 0.126}$&$0.732^{\pm 0.009}$&36.87/41 \\ 
 9&X-02-00&53560.51&$1.315^{\pm 0.079}$&$1.646^{\pm 0.013}$&$0.362^{\pm 0.079}$&$7.343^{\pm 0.035}$&$52.72^{}$&$5.61^{\pm 0.36}$&$2.080^{\pm 0.073}$&$0.890^{\pm 0.030}$&$0.428^{\pm 0.029}$&$48.73^{\pm  3.77}$&$1.097^{\pm 0.053}$&$1.632^{\pm 0.117}$&$0.720^{\pm 0.007}$&26.49/41 \\ 
10&Y-02-00&53563.08&$1.313^{\pm 0.083}$&$1.611^{\pm 0.012}$&$0.317^{\pm 0.083}$&$6.813^{\pm 0.306}$&$50.17^{}$&$5.52^{\pm 0.31}$&$1.959^{\pm 0.041}$&$0.865^{\pm 0.033}$&$0.441^{\pm 0.026}$&$41.00^{\pm  3.93}$&$1.095^{\pm 0.060}$&$2.315^{\pm 0.244}$&$0.608^{\pm 0.008}$&36.92/41 \\ 
11&Y-02-05&53563.53&$1.281^{\pm 0.096}$&$1.626^{\pm 0.012}$&$0.265^{\pm 0.096}$&$6.761^{\pm 0.030}$&$49.60^{}$&$5.59^{\pm 0.36}$&$1.910^{\pm 0.042}$&$0.848^{\pm 0.030}$&$0.444^{\pm 0.026}$&$42.38^{\pm  2.03}$&$1.100^{\pm 0.035}$&$2.332^{\pm 0.175}$&$0.579^{\pm 0.008}$&37.69/40 \\ 
12&Y-02-06&53564.91&$1.298^{\pm 0.092}$&$1.609^{\pm 0.012}$&$0.271^{\pm 0.092}$&$6.563^{\pm 0.029}$&$48.21^{}$&$5.45^{\pm 0.32}$&$1.824^{\pm 0.056}$&$0.829^{\pm 0.026}$&$0.454^{\pm 0.028}$&$52.14^{\pm  4.03}$&$1.084^{\pm 0.042}$&$2.615^{\pm 0.218}$&$0.582^{\pm 0.008}$&21.55/41 \\ 
13&Y-03-00&53567.09&$1.316^{\pm 0.092}$&$1.583^{\pm 0.012}$&$0.241^{\pm 0.092}$&$5.982^{\pm 0.236}$&$47.13^{}$&$5.46^{\pm 0.27}$&$1.760^{\pm 0.056}$&$0.804^{\pm 0.027}$&$0.457^{\pm 0.030}$&$52.82^{\pm  2.99}$&$1.093^{\pm 0.069}$&$2.072^{\pm 0.241}$&$0.485^{\pm 0.006}$&34.16/41 \\ 
14&Y-03-02&53568.58&$1.427^{\pm 0.090}$&$1.583^{\pm 0.015}$&$0.274^{\pm 0.090}$&$5.759^{\pm 0.031}$&$42.91^{}$&$5.48^{\pm 0.26}$&$1.672^{\pm 0.043}$&$0.779^{\pm 0.027}$&$0.466^{\pm 0.028}$&$51.84^{\pm  3.06}$&$1.073^{\pm 0.039}$&$1.750^{\pm 0.133}$&$0.511^{\pm 0.011}$&53.84/41 \\ 
15&Y-03-03&53569.75&$1.462^{\pm 0.085}$&$1.583^{\pm 0.016}$&$0.312^{\pm 0.084}$&$5.813^{\pm 0.035}$&$37.45^{}$&$5.46^{\pm 0.27}$&$1.611^{\pm 0.049}$&$0.745^{\pm 0.028}$&$0.462^{\pm 0.031}$&$48.71^{\pm  4.02}$&$1.086^{\pm 0.030}$&$1.797^{\pm 0.130}$&$0.570^{\pm 0.014}$&36.33/41 \\ 
16&Y-03-04&53570.54&$1.417^{\pm 0.094}$&$1.581^{\pm 0.014}$&$0.251^{\pm 0.093}$&$5.530^{\pm 0.029}$&$37.69^{}$&$5.08^{\pm 0.22}$&$1.559^{\pm 0.045}$&$0.729^{\pm 0.029}$&$0.468^{\pm 0.032}$&$58.34^{\pm  5.40}$&$1.094^{\pm 0.050}$&$1.652^{\pm 0.133}$&$0.506^{\pm 0.010}$&62.48/40 \\ 
17&Y-03-05&53571.53&$1.385^{\pm 0.088}$&$1.569^{\pm 0.014}$&$0.256^{\pm 0.087}$&$5.331^{\pm 0.255}$&$36.46^{}$&$5.02^{\pm 0.17}$&$1.505^{\pm 0.049}$&$0.698^{\pm 0.026}$&$0.464^{\pm 0.032}$&$54.17^{\pm  4.18}$&$1.119^{\pm 0.053}$&$1.628^{\pm 0.126}$&$0.492^{\pm 0.011}$&62.74/40 \\ 
18&Y-03-06&53572.91&$1.311^{\pm 0.102}$&$1.579^{\pm 0.013}$&$0.197^{\pm 0.102}$&$5.097^{\pm 0.042}$&$38.98^{}$&$5.22^{\pm 0.23}$&$1.411^{\pm 0.055}$&$0.669^{\pm 0.025}$&$0.474^{\pm 0.036}$&$60.50^{\pm  3.16}$&$1.101^{\pm 0.073}$&$1.550^{\pm 0.129}$&$0.462^{\pm 0.008}$&72.73/40 \\ 
19&Y-04-00&53573.89&$1.409^{\pm 0.087}$&$1.557^{\pm 0.015}$&$0.242^{\pm 0.086}$&$4.845^{\pm 0.240}$&$39.20^{}$&$5.12^{\pm 0.23}$&$1.356^{\pm 0.043}$&$0.630^{\pm 0.024}$&$0.464^{\pm 0.032}$&$52.23^{\pm  5.37}$&$1.130^{\pm 0.090}$&$1.543^{\pm 0.130}$&$0.461^{\pm 0.007}$&52.38/40 \\ 
20&Y-04-01&53574.68&$1.418^{\pm 0.097}$&$1.565^{\pm 0.015}$&$0.217^{\pm 0.096}$&$4.845^{\pm 0.212}$&$36.29^{}$&$5.43^{\pm 0.26}$&$1.330^{\pm 0.043}$&$0.630^{\pm 0.025}$&$0.474^{\pm 0.034}$&$61.45^{\pm  4.36}$&$1.096^{\pm 0.083}$&$1.595^{\pm 0.135}$&$0.458^{\pm 0.010}$&61.22/40 \\ 
21&Y-04-06&53579.45&$1.452^{\pm 0.090}$&$1.566^{\pm 0.016}$&$0.193^{\pm 0.089}$&$4.332^{\pm 0.024}$&$33.84^{}$&$5.42^{\pm 0.29}$&$1.130^{\pm 0.046}$&$0.568^{\pm 0.024}$&$0.503^{\pm 0.042}$&$71.67^{\pm  5.74}$&$1.101^{\pm 0.097}$&$1.668^{\pm 0.123}$&$0.473^{\pm 0.011}$&67.70/41 \\ 
22&Y-05-01&53583.66&$1.531^{\pm 0.102}$&$1.528^{\pm 0.016}$&$0.185^{\pm 0.007}$&$3.719^{\pm 0.022}$&$29.54^{}$&$5.84^{\pm 0.12}$&$1.108^{\pm 0.044}$&$0.550^{\pm 0.021}$&$0.507^{\pm 0.010}$&$70.14^{\pm  5.77}$&$1.135^{\pm 0.092}$&$1.578^{\pm 0.115}$&$0.365^{\pm 0.006}$&64.73/40  \\
23&Y-06-00&53587.65&$1.634^{\pm 0.097}$&$1.503^{\pm 0.017}$&$0.168^{\pm 0.011}$&$3.228^{\pm 0.082}$&$26.00^{}$&$5.90^{\pm 0.11}$&$1.020^{\pm 0.045}$&$0.518^{\pm 0.022}$&$0.512^{\pm 0.010}$&$81.58^{\pm  5.23}$&$1.218^{\pm 0.088}$&$1.613^{\pm 0.113}$&$0.294^{\pm 0.005}$&74.74/41 \\
  & & & & & & & & & & & & & & & \\
24&Y-06-01&53589.49&$1.536^{\pm 0.120}$&$1.516^{\pm 0.017}$&$0.111^{\pm 0.119}$&$2.934^{\pm 0.099}$&$21.20^{}$&$4.76^{\pm 0.27}$&$0.989^{\pm 0.045}$&$0.507^{\pm 0.020}$&$0.513^{\pm 0.044}$&$86.18^{\pm  7.70}$&$1.745^{\pm 0.097}$&$1.500^{\pm 0.125}$&$0.242^{\pm 0.009}$&45.95/40 \\ 
25&Y-06-05&53590.81&$1.723^{\pm 0.110}$&$1.493^{\pm 0.015}$&$0.131^{\pm 0.109}$&$2.877^{\pm 0.026}$&$19.92^{}$&$4.75^{\pm 0.29}$&$0.960^{\pm 0.046}$&$0.497^{\pm 0.021}$&$0.517^{\pm 0.047}$&$93.63^{\pm  8.59}$&$1.817^{\pm 0.100}$&$1.550^{\pm 0.138}$&$0.225^{\pm 0.007}$&45.84/40 \\ 
26&Y-06-06&53591.79&$1.601^{\pm 0.113}$&$1.495^{\pm 0.016}$&$0.121^{\pm 0.112}$&$2.722^{\pm 0.019}$&$16.53^{}$&$5.33^{\pm 0.23}$&$0.904^{\pm 0.047}$&$0.460^{\pm 0.023}$&$0.508^{\pm 0.051}$&$109.5^{\pm  9.12}$&$1.832^{\pm 0.149}$&$1.521^{\pm 0.121}$&$0.216^{\pm 0.006}$&48.48/40 \\ 
27&Y-06-07&53591.92&$1.417^{\pm 0.099}$&$1.510^{\pm 0.019}$&$0.173^{\pm 0.098}$&$2.845^{\pm 0.025}$&$18.85^{}$&$5.52^{\pm 0.20}$&$0.863^{\pm 0.042}$&$0.441^{\pm 0.019}$&$0.510^{\pm 0.047}$&$137.4^{\pm 11.87}$&$1.918^{\pm 0.145}$&$1.646^{\pm 0.122}$&$ - - - -         $&37.40/41 \\ 
28&Y-06-03&53593.23&$1.530^{\pm 0.133}$&$1.516^{\pm 0.020}$&$0.103^{\pm 0.132}$&$2.689^{\pm 0.021}$&$16.61^{}$&$5.50^{\pm 0.26}$&$0.825^{\pm 0.044}$&$0.421^{\pm 0.019}$&$0.511^{\pm 0.050}$&$137.3^{\pm 14.04}$&$1.936^{\pm 0.105}$&$1.790^{\pm 0.131}$&$0.237^{\pm 0.005}$&50.79/41 \\ 
29&Y-07-00&53595.73&$1.651^{\pm 0.164}$&$1.464^{\pm 0.033}$&$0.149^{\pm 0.164}$&$2.446^{\pm 0.042}$&$19.43^{}$&$5.43^{\pm 0.21}$&$0.817^{\pm 0.035}$&$0.417^{\pm 0.017}$&$0.510^{\pm 0.043}$&$137.7^{\pm 13.99}$&$1.933^{\pm 0.102}$&$1.736^{\pm 0.128}$&$0.211^{\pm 0.006}$&67.79/41 \\ 
30&Y-07-01&53597.30&-- --&$1.587^{\pm 0.024}$&-- --&$2.466^{\pm 0.042}$&-- --&$5.50^{\pm 0.27}$&$0.775^{\pm 0.030}$&$0.399^{\pm 0.016}$&$0.515^{\pm 0.040}$&$141.9^{\pm 14.15}$&$2.115^{\pm 0.156}$&$1.432^{\pm 0.115}$&$0.232^{\pm 0.008}$&46.49/41 \\ 
31&Y-08-01&53603.38&-- --&$1.586^{\pm 0.034}$&-- --&$2.168^{\pm 0.062}$&-- --&$5.09^{\pm 0.21}$&$0.730^{\pm 0.022}$&$0.378^{\pm 0.012}$&$0.518^{\pm 0.032}$&$162.6^{\pm 14.50}$&$2.893^{\pm 0.167}$&$1.600^{\pm 0.117}$&$0.197^{\pm 0.006}$&53.12/43 \\ 
32&Y-08-02&53605.47&-- --&$1.586^{\pm 0.022}$&-- --&$2.055^{\pm 0.025}$&-- --&$5.01^{\pm 0.33}$&$0.659^{\pm 0.030}$&$0.350^{\pm 0.014}$&$0.530^{\pm 0.045}$&$215.7^{\pm 18.83}$&$2.981^{\pm 0.172}$&$1.533^{\pm 0.119}$&$ - - - -         $&33.77/40 \\ 
33&Y-08-03&53607.25&-- --&$1.587^{\pm 0.031}$&-- --&$1.998^{\pm 0.035}$&-- --&$4.98^{\pm 0.30}$&$0.632^{\pm 0.029}$&$0.331^{\pm 0.008}$&$0.522^{\pm 0.037}$&$242.0^{\pm 17.71}$&$3.126^{\pm 0.161}$&$1.846^{\pm 0.123}$&$ - - - -         $&44.67/43 \\ 
34&Y-09-00&53612.75&-- --&$1.583^{\pm 0.021}$&-- --&$1.917^{\pm 0.030}$&-- --&$5.47^{\pm 0.25}$&$0.622^{\pm 0.029}$&$0.329^{\pm 0.006}$&$0.529^{\pm 0.035}$&$242.2^{\pm 16.29}$&$3.101^{\pm 0.162}$&$1.658^{\pm 0.126}$&$0.190^{\pm 0.005}$&56.21/41 \\ 
35&Y-10-00&53619.69&-- --&$1.581^{\pm 0.032}$&-- --&$1.687^{\pm 0.025}$&-- --&$5.48^{\pm 0.26}$&$0.624^{\pm 0.015}$&$0.325^{\pm 0.007}$&$0.521^{\pm 0.024}$&$268.0^{\pm 22.07}$&$2.983^{\pm 0.164}$&$1.613^{\pm 0.123}$&$ - - - -         $&34.92/41 \\ 
36&Y-11-00&53623.76&-- --&$1.586^{\pm 0.022}$&-- --&$1.527^{\pm 0.023}$&-- --&$5.23^{\pm 0.28}$&$0.579^{\pm 0.018}$&$0.304^{\pm 0.010}$&$0.525^{\pm 0.033}$&$268.1^{\pm 23.50}$&$3.103^{\pm 0.164}$&$1.638^{\pm 0.122}$&$ - - - -         $&48.11/40 \\ 
37&Y-12-00&53630.31&-- --&$1.578^{\pm 0.043}$&-- --&$1.236^{\pm 0.047}$&-- --&$5.34^{\pm 0.30}$&$0.570^{\pm 0.017}$&$0.294^{\pm 0.009}$&$0.516^{\pm 0.032}$&$268.0^{\pm 21.16}$&$3.218^{\pm 0.161}$&$1.414^{\pm 0.114}$&$0.193^{\pm 0.006}$&43.00/44 \\ 
38&Y-12-01&53634.24&-- --&$1.578^{\pm 0.027}$&-- --&$1.167^{\pm 0.022}$&-- --&$5.21^{\pm 0.22}$&$0.564^{\pm 0.018}$&$0.294^{\pm 0.009}$&$0.520^{\pm 0.033}$&$247.3^{\pm 16.73}$&$3.340^{\pm 0.175}$&$1.484^{\pm 0.121}$&$ - - - -         $&64.99/44 \\ 
39&Y-13-00&53638.65&-- --&$1.571^{\pm 0.033}$&-- --&$1.085^{\pm 0.029}$&-- --&$4.96^{\pm 0.29}$&$0.563^{\pm 0.015}$&$0.289^{\pm 0.010}$&$0.513^{\pm 0.031}$&$275.9^{\pm 18.50}$&$3.113^{\pm 0.197}$&$1.520^{\pm 0.123}$&$ - - - -         $&69.31/44 \\ 
40&Y-13-01&53641.60&-- --&$1.581^{\pm 0.030}$&-- --&$1.075^{\pm 0.029}$&-- --&$5.03^{\pm 0.27}$&$0.553^{\pm 0.019}$&$0.287^{\pm 0.009}$&$0.519^{\pm 0.033}$&$270.7^{\pm 18.13}$&$3.277^{\pm 0.197}$&$1.497^{\pm 0.111}$&$ - - - -         $&66.99/43 \\ 
41&Y-14-00&53646.98&-- --&$1.580^{\pm 0.031}$&-- --&$1.019^{\pm 0.020}$&-- --&$4.95^{\pm 0.28}$&$0.549^{\pm 0.019}$&$0.282^{\pm 0.010}$&$0.513^{\pm 0.036}$&$278.9^{\pm 18.36}$&$3.189^{\pm 0.182}$&$1.539^{\pm 0.121}$&$ - - - -         $&63.53/44 \\ 
42&Y-15-00&53652.67&-- --&$1.532^{\pm 0.022}$&-- --&$1.000^{\pm 0.019}$&-- --&$5.45^{\pm 0.29}$&$0.539^{\pm 0.018}$&$0.281^{\pm 0.008}$&$0.520^{\pm 0.032}$&$290.5^{\pm 19.93}$&$3.382^{\pm 0.205}$&$1.450^{\pm 0.137}$&$ - - - -         $&58.29/44 \\ 
43&Y-15-01&53654.63&-- --&$1.595^{\pm 0.032}$&-- --&$0.997^{\pm 0.020}$&-- --&$5.41^{\pm 0.17}$&$0.516^{\pm 0.019}$&$0.275^{\pm 0.010}$&$0.533^{\pm 0.039}$&$282.9^{\pm 18.96}$&$3.366^{\pm 0.278}$&$1.450^{\pm 0.136}$&$ - - - -         $&49.88/44 \\ 
44&Y-16-01&53662.77&-- --&$1.591^{\pm 0.031}$&-- --&$0.987^{\pm 0.018}$&-- --&$5.57^{\pm 0.16}$&$0.497^{\pm 0.017}$&$0.269^{\pm 0.010}$&$0.541^{\pm 0.037}$&$279.7^{\pm 18.89}$&$3.427^{\pm 0.230}$&$1.480^{\pm 0.122}$&$ - - - -         $&63.57/44 \\ 

\hline
\end{tabular}
\noindent{
\leftline {X=91094-01 and Y=91423-01 are the prefixes of observation Ids. Blank lines mark transitions between different spectral states.}
\leftline {$T_{in}$, and $\Gamma$ values indicate combined DBB and PL model fitted DBB temperatures in keV and PL photon indices respectively. }
\leftline {$^\dagger$ DBBf, PLf represent combined DBB and PL model fitted fluxes in 2.5-25~keV band for DBB and PL model components respectively in units of $10^{-9}~ergs~cm^{-2}~s^{-1}$. }
\leftline {$\dot{m_h}$ (in Eddington rate unit), $\dot{m_d}$ (in Eddington rate unit), $X_s$ (in units of Schwarzschild radius), and $R$ are TCAF fitted Keplerian disk, sub-Keplerian halo,} 
\leftline {shock location and compression ratio values respectively. Norm. and $M_{BH}$ are TCAF fitted normalization and BH mass values respectively.} 
\leftline {$^{\dagger\dagger}$ Frequencies of the dominating QPO in Hz are mentioned. DOF means degrees of freedom of the spectral model fits.}
\leftline {The values of $\chi^2$, and DOF of TCAF model fitted spectra are mentioned in Col. 17. The ratio is the $\chi^2_{red.}$.}
\leftline {Note: average values of 90\% confidence $\pm$ error values obtained using `err' task in XSPEC, are placed as superscripts of fitted parameters.}
\leftline {Also note that the accretion rate ${\dot m}_d$ is sometimes higher than the Eddington rate. Even though the luminosity may be sub-Eddington. }
\leftline {This is because TCAF self-consistently computes the efficiency factor $\eta <1$ for each data and the observed luminosity $L=\eta L_E$.}
}
\end{table}

\newpage
\clearpage

\end{document}